\documentclass[12pt]{article}
\usepackage{authblk}
\usepackage{verbatim}
\usepackage{caption}
\usepackage{adjustbox}
\usepackage{amsfonts}
\usepackage{graphics}
\usepackage{amsmath}
\usepackage{times}
\usepackage{appendix}
\usepackage{color}
\usepackage{mathtools}
\usepackage{enumerate}
\usepackage{fancyhdr,latexsym,amsmath,amsfonts,amssymb,amsbsy,amsthm,url}
\usepackage[margin=1in,footskip=0.5in]{geometry}
\usepackage{graphics,graphicx,epsfig}
\usepackage{breqn}
\usepackage{caption,subcaption,float,subfloat}
\usepackage{pdflscape}
\usepackage{hyperref}
\usepackage[ruled,vlined]{algorithm2e}
\usepackage{enumitem,amssymb}
\newlist{todolist}{itemize}{2}
\setlist[todolist]{label=$\square$}
\newtheorem{theorem}{Theorem}[]

\usepackage[english]{babel}
\usepackage[dvipsnames]{xcolor}
\makeatletter
\providecommand{\keywords}[1]
{
  \small	
  \textit{Keywords:} #1
}

\renewcommand{\section}{
	\@startsection
	{section}% name
	{1}% level
	{0pt}% indent
	{1.1\baselineskip}% beforeskip
	{0.2\baselineskip}% afterskip
	{\sc \centering}% style
}

\renewcommand{\subsection}{
	\@startsection
	{subsection}% name
	{1}% level
	{0pt}% indent
	{1.1\baselineskip}% beforeskip
	{0.2\baselineskip}% afterskip
	{\sc \centering}% style
}

\renewcommand{\subsubsection}{
	\@startsection
	{subsubsection}% name
	{1}% level
	{0pt}% indent
	{1.1\baselineskip}% beforeskip
	{0.2\baselineskip}% afterskip
	{\sc \centering}% style
}

\makeatother

\newcommand{\innerproduct}[2]{\langle #1, #2 \rangle}

\newcommand{\abs}[1]{\lvert #1 \rvert}
\newcommand{\babs}[1]{\bigg \lvert #1 \bigg \rvert}
\usepackage[flushleft]{threeparttable}
\usepackage{rotating,booktabs,multirow}
\usepackage{colortbl}
\usepackage{makecell,cellspace,caption}

\begin{document}
	
\title{\large\sc Multilevel Monte Carlo and its Applications in Financial Engineering}
\normalsize
\author[1]{Devang Sinha}
\author[2]{Siddhartha P. Chakrabarty}
\affil[1]{Department of Mathematics, Indian Institute of Technology Guwahati, Guwahati-781039, Assam, India, e-mail: dsinha@iitg.ac.in}
\affil[2]{Department of Mathematics, Indian Institute of Technology Guwahati, Guwahati-781039, Assam, India, e-mail: pratim@iitg.ac.in,
Phone: +91-361-2582606, Fax: +91-361-2582649}
\date{}
\maketitle

\begin{abstract}
	
In this article, we present a review of the recent developments on the topic of Multilevel Monte Carlo (MLMC) algorithm, in the paradigm of applications in financial engineering. We specifically focus on the recent studies conducted in two subareas, namely, option pricing and financial risk management. For the former, the discussion involves incorporation of the importance sampling algorithm, in conjunction with the MLMC estimator, thereby constructing a hybrid algorithm in order to achieve reduction for  the overall variance of the estimator. In case of the latter, we discuss the studies carried out in order to construct an efficient algorithm in order to estimate the risk measures of Value-at-Risk (VaR) and Conditional Var (CVaR), in an efficient manner. In this regard, we briefly discuss the motivation and the construction of an adaptive sampling algorithm with an aim to efficiently estimate the nested expectation, which, in general is computationally expensive.
\end{abstract}

\begin{itemize}
\item Brief overview of the Multilevel Monte Carlo estimator.
\item Review of the importance sampling algorithm to reduce the overall variance of the MLMC estimator associated with the option pricing problems in financial engineering.
\item Review of extension of adaptive sampling algorithm to multilevel paradigm, with the aim of improving the computational complexity while estimating the Value-at-Risk (VaR) and Conditional VaR (CVaR).
\end{itemize}

\begin{table}[]
\centering
\begin{adjustbox}{width=\columnwidth,center}
\begin{tabular}{|l|l|}
\hline
\textbf{Subject area}                     & Mathematics and Statistics                                                                      \\ \hline
\textbf{More specific subject area}       & Computational Finance                                                                           \\ \hline
\textbf{Name of the reviewed methodology} & Importance Sampling for Option Pricing and Adaptive Sampling in Efficient Risk Estimation.      \\ \hline
\textbf{Keywords}                         & MLMC; Option Pricing; Importance Sampling; Risk Management; Adaptive sampling \\ \hline
\textbf{Resource availability}            & N/A                                                                                             \\ \hline
\textbf{Review question} &
\begin{tabular}[c]{@{}l@{}}
1. What are recent development in the field of Multilevel Monte Carlo?\\ 
2. How have these developments led to the improved efficiency of the existing estimator under various financial applications?\\ 
3. What are the shortcomings of the presented studies?\\ 
4. What are the recent developments catering to these shortcoming?\end{tabular} \\ 
\hline
\end{tabular}
\end{adjustbox}
\caption{Specifications Table}
\label{Specifications table}
\end{table}

\section{Introduction}
\label{intro}

In the broader area of computational finance, the mere establishment of the existence of solution to a problem is not sufficient towards achieving the tangible financial goals, for the problem that has been posed. Accordingly, as is the case for many applications, we seek a solution that (in practice) happens to be an approximation to the actual solution being sought. To this end, we begin by observing that for problems in quantitative finance, one can arrive at either an analytical or (possibly) a semi-analytical solution, in only a handful of cases. Therefore, for  
the most part, one needs to devise efficient methods to arrive at the desired and appropriate solution to the posed problem which, in turn, necessitates the resorting to computational techniques. At the heart of this paper, lies a specific computational technique, widely used in the finance industry, namely, the Monte Carlo simulation approach. Accordingly, we begin our presentation with a brief narrative about this approach, with the main focus of the discussion being steered towards the
recent research and development in the area of Multilevel Monet Carlo (MLMC). The interested readers may refer to \cite{giles2008multilevel,giles2008improved,giles2013antithetic,giles2013multilevel,lemaire2017multilevel,belomestny2013multilevel,heinrich2001multilevel} for greater clarity on the approach of MLMC and the key developments with respect to algorithms, as well as the applications in financial engineering problems. In this article we primarily focus on the importance sampling approach developed and studied in \cite{kebaier2018coupling,alaya2016improved} and also on how MLMC has led to the development of algorithms for efficient risk estimation in the field of financial risk management, discussed by the authors in \cite{giles2019multilevel}. However, we give a brief overview of Monte Carlo and MLMC before directing our discussion towards the aforesaid specific topics.

Monte Carlo methods have become one of the driving computing tools in the finance industry. The necessity of simulating high-dimensional stochastic models, which in turn may be attributed to the linear development in the complexity corresponding to the size of the problem itself, is one of the primary reasons that this approach is becoming the critical computational strategy in the industry. The main objective of this method, in case of computational finance is to reach the necessary degree of accuracy, which is coupled with a high computational cost. More specifically, we intend to approximate $\mathbf{E}[Y]$ where, $Y = G(X)$ is functional of the random variable $X$. The traditional Monte Carlo approach requires a computational complexity of an order of $O(\epsilon^{-3})$ to attain the root mean square (RMS) error of $O(\epsilon)$ in a biased context \cite{giles2008multilevel}. This limitation led to the introduction of the multilevel framework in \cite{giles2008multilevel} to address this issue and achieve $O(\epsilon^{-2})$ computational complexity in the biased framework.

The idea behind the multilevel architecture is to employ independent standard Monte Carlo on various resolution levels and use the differences as the control variate for the Monte Carlo simulation at its most granular level, which in mathematical terms is given by,
\begin{equation}
\label{rev:eq1}
\mathbf{E}[Y_{L}]=\mathbf{E}[Y_{1}]+\sum\limits_{l=2}^{L}\mathbf{E}[Y_{l}-Y_{l-1}],~\text{where}~ Y_{l}=G(X_{l}).
\end{equation}
Using the standard Monte Carlo as the estimator to approximate the expectation on the right hand side of \eqref{rev:eq1}, we obtain,
\begin{equation}
\label{rev:eq2}
\widehat{Y}_{L}=\frac{1}{N_{1}}\sum\limits_{k=1}^{N_{1}}Y_{1}^{k}+\sum\limits_{l=2}^{L}\frac{1}{N_{l}}\sum\limits_{k=1}^{N_{l}}\left(Y_{l}^{k}-Y_{l-1}^{k}\right),
\end{equation}
and therefore, $\mathbf{E}[Y_{L}] \approx \widehat{Y}_{L}$. Here $X_{l}$ is the approximation of the random variable $X$ on level $l$ and this approximation is contingent on the application under consideration. For example, if the underlying stochastic process is driven by a stochastic differential equation (SDE), then $X_{l}$ is the approximation of $X$, with some time discretization parameter $h_{l}$. With all the preludes being presented in the preceding discussion, we are now in a position to examine the following the seminal result due to Giles \cite{giles2008multilevel}.
\begin{theorem}
Let $G$ denote a functional of the random variable X, and let $Y_{l} = G(X_l)$ denote the corresponding level $l$ numerical approximation. If there exist independent estimators $Z_{l}$, based on $N_{l}$ Monte Carlo samples, and positive constants $\alpha,\beta,\gamma, c_{1},c_{2},c_{3},c_{4}$ such that $\displaystyle{\alpha \geq \frac{1}{2}\min\left(\alpha,\beta\right)}$ and
\begin{enumerate}
\item $\displaystyle{\left|\mathbf{E}\left[Y_{l}-Y_{l-1}\right]\right| \leq c_{1}2^{-\alpha l}}$.
\item $\displaystyle{\mathbf{E}\left[Z_{l}\right]=
\begin{cases}
\mathbf{E}\left[Y_{1}\right],~l=0, \\
\mathbf{E}\left[Y_{l}-Y_{l-1}\right],~l>0.
\end{cases}}$.
\item $\displaystyle{\mathbf{V}[Z_{l}] \leq c_{2}N_{l}^{-1}2^{-\beta l}}$.
\item $\displaystyle{C_{l} \leq c_{3}N_{l}2^{\gamma l}}$, where $C_{l}$ is the computational complexity of $Z_{l}$,
\end{enumerate}
then there exists a positive constant $c_{4}$ such that for any $\epsilon < e^{-1}$, there are values $L$ and $N_{l}$ for which the multilevel estimator $\displaystyle{Z=\sum\limits_{l=1}^{L} Z_{l}}$,has a MSE with bound,
\[MSE\equiv \mathbf{E}\left[\left(Z-\mathbf{E}[Y]\right)^{2}\right] < \epsilon^{2},\]
with a computational complexity $C$, having the bound,
\[\mathbf{E}[C] \leq 
\begin{cases}
c_{4}\epsilon^{-2},~\beta>\gamma, \\
c_{4}\epsilon^{-2}\left(\log\epsilon\right)^{2},~\beta=\gamma, \\
c_{4}\epsilon^{-2-\frac{(\gamma-\beta)}{\alpha}},~0<\beta <\gamma.
\end{cases}\]
\end{theorem}
It is quite evident from the above theorem that the computational complexity is driven by the strong convergence of the estimator \textit{i.e.,} $\mathbf{V}[Z_{l}]$. Therefore, one of the main challenges while developing a MLMC based estimator is to study the order of strong convergence of the underlying approximation.  With this brief introduction of MLMC, we now direct our discussion towards its recent developments, pertaining to algorithm and financial applications.

\section{Importance Sampling Multilevel Algorithm}
\label{importance_sampling}

Since the advent of MLMC in literature, one of the directions of its progression has been through various attempts to combine this algorithm, with the already existing variance reduction techniques. For instance, Giles, in \cite{giles2013antithetic, giles2014antithetic} studied and analyzed the combination of antithetic variates and MLMC in order to bypass the Levy area simulation, encountered while using Milstein discretization scheme, in order to simulate higher dimensional SDEs. However, in our discussion we primarily focus on the combination of importance sampling algorithm and Multilevel estimator.

The idea of incorporating importance sampling with multilevel estimators is derived from the seminal paper by Arouna \cite{arouna2004adaptative}. Arouna's idea relied upon the parametric change of measure and using a search algorithm to approximate the optimal change of the measure parameter, in order to minimize the variance of the standard Monte Carlo estimator. Before we discuss the research undertaken in the area of multilevel pertaining to importance sampling algorithm, we give a brief overview of the parametric importance sampling approach.

Consider a general problem of estimating $\mathbf{E}[G(X)]$, where $X$ is a $d$-dimensional random variable. If $f(x)$ is the multivariate density function, then,
\[\mathbf{E}[G(X)]=\int G(x)f(x)dx=\int G(x+\theta)f(x+\theta)dx=\int h(\theta, x)f(x)dx,\] 
where,
$\displaystyle{h(\theta,x)=\frac{G(x+\theta)f(x+\theta)}{f(x)}}$.
This implies that, $\mathbf{E}[G(X)]=\mathbf{E}[h(\theta,X)]$. Therefore, we need to determine the optimal value of $\theta$ such that $\text{Var}[h(\theta,X)]$ is minimum. Mathematically this is represented as,
\begin{equation}
\label{rev:eq3}
\theta^{*}=\arg\min_{\theta \in \mathbf{R}^{d}}\text{Var}[h(\theta,X)].
\end{equation}
In order to solve the above problem, one can resort to the usage of the Robbins-Monro algorithm that deals with a sequence of random variable $\left(\theta_{i}\right)_{i\in\mathbb{N}}$, which approximate $\theta^{*}$ accurately. However, the convergence of this algorithm requires certain restrictive conditions, which are known as the non explosion condition (given in \cite{alaya2015importance}),
\[\mathbf{E}[h^{2}\left(\theta, X\right)] \leq C\left(1+|\theta|^{2}\right)~\text{for all}~\theta \in \mathbf{R}^{d}.\]

In order to deal with this restrictive condition, the authors in \cite{chen1987convergence,chen1986stochastic} introduced a truncation based procedure which was furthered in \cite{andrieu2005stability,lelong2008almost}. An unconstrained procedure to approximate $\theta^{*}$, by using the regularity of the density function in an extensive manner, was introduced in \cite{lemaire2010unconstrained} along with the proof of convergence of the algorithm. Beside the stochastic approximation algorithm, one can also use deterministic algorithm such as sample average approximation, which, though being computationally expensive, provides for a better approximation to $\theta^{*}$.

In problems dealing with the pricing of the options, for the most part the underlying stochastic process $(X_t)_{0\leq t \leq T}$ with $T>0$ being a finite time horizon, is governed by some SDEs. The general form of these SDEs is given as follows:
\begin{equation}
\label{rev:eq4}
dX_{t}=b(X_{t})dt+\sum\limits_{j=1}^{q} \sigma_{j}(X_{t})dW_{t}^{j},~X_{0}=x \in \mathbb{R}^{d}, 
\end{equation}
where, 
$\displaystyle{W:=\begin{pmatrix}W_{1}&W_{2}&\dots&W_{q}\end{pmatrix}}$ is a $q$-dimensional Brownian motion on a filtered probability space $\left(\Omega,(\mathcal{F}_{t}\right)_{0 \leq t \leq T},\mathbb{P})$, with $b:\mathbb{R}^{d} \rightarrow \mathbb{R}^{d}$ and $\sigma_{j}:\mathbb{R}^{d} \rightarrow \mathbb{R}^{d}$ being the functions satisfying the following condition:
\begin{equation}
\label{rev:assumption_1}
\forall x,y \in \mathbb{R}^{d},~\lvert b(x)-b(y)\rvert+\sum\limits_{j=1}^{q}\lvert \sigma_{j}(x)-\sigma_{j}(y)\rvert  < K_{b,\sigma}\lvert x-y \rvert,~\text{where}~ K_{b,\sigma}>0. 
\tag{$\text{C}_1$}
\end{equation}
Assumption \eqref{rev:assumption_1} ensures the existence and the uniqueness of solution to \eqref{rev:eq4}. For the most part, constructing an analytical or semi-analytical solution to \eqref{rev:eq4} is not possible, and therefore we need to rely on discretization schemes such as Euler or Milstein in order to simulate the SDEs. For detailed discussion on these discretization schemes, the interested readers may refer to \cite{kloeden1992stochastic}. Further, following the idea of \cite{arouna2004adaptative}, we consider a family of stochastic process $\left(X_{t}(\theta)\right)_{0 \leq t \leq T}$, with $\theta \in \mathbb{R}^{d}$, being governed by the following SDE:
\begin{equation}
\label{rev:eq5}
dX_{t}(\theta) = (b(X_{t}(\theta))+\sigma(X_{t}(\theta))\theta)dt + \sum\limits_{j=1}^{q}\sigma_{j}(X_{t}(\theta))dW_{t}^{j},~\sigma(x)=\begin{pmatrix}\sigma_{1}(x)&\dots&\sigma_{q}(x)\end{pmatrix}.
\end{equation}
As a consequence of the Girsanov's Theorem, we know that there exists a risk-neutral probability measure $\mathbf{P}_{\theta}$, which is equivalent to $\mathbf{P}$ such that,
\begin{equation}
\label{rev:eq6}
\frac{d\mathbf{P}_{\theta}}{{d\mathbf{P}|}_{\mathcal{F}_t}}=\exp{\left(-\innerproduct{\theta}{ W_{t}}- \frac{1}{2}\lvert\theta\rvert^{2}t\right)},
\end{equation}
under which the process $\displaystyle{\left(\theta t+W_{t}\right)_{0\leq t \leq T}}$ is a Brownian motion. Therefore,
\begin{equation}
\label{rev:eq7}
\mathbf{E}_{\mathbf{P}}\left[G(X_{T})\right] = \mathbf{E}_{\mathbf{P}_{\theta}}\left[G(X_{T}(\theta))\right]= \mathbf{E}_{\mathbf{P}}\left[G(X_{T}(\theta))e^{-\innerproduct{\theta}{ W_{T}}- \frac{1}{2}\lvert\theta\rvert^{2}T}\right]
\end{equation}
Therefore, following the discussion above we have, 
\[\mathbf{E}\left[G(X_{T})\right]=\mathbf{E}\left[h(\theta, X_{T})\right].\]
here, $h(\theta, X_{T}) = G(X_T(\theta))e^{-\innerproduct{\theta}{ W_{T}}- \frac{1}{2}\lvert\theta\rvert^{2}T}$.
Now the idea of importance sampling Monte Carlo method is to estimate $\mathbf{E}\left(G(X_{T})\right)$, where $\theta$ is given by,
\begin{equation}
\label{rev:eq8}
\theta^{*}=\arg\min_{\theta \in \mathbb{R}^{d}} \text{Var}~\left(G(X_{T}(\theta))e^{-\innerproduct{\theta}{ W_{T}}- \frac{1}{2}\lvert\theta\rvert^{2}T}\right).  
\end{equation}
In the context of Multilevel estimator, we present two approaches studied in \cite{kebaier2018coupling,alaya2016improved}, adapting the ideas studied by authors in \cite{arouna2004adaptative,alaya2015importance} and extending it to multilevel scenarios. Under the parametric change of measure, the general multilevel estimator is given defines as,
\begin{equation}
\label{rev:eq9}
\mathbf{E}[Y_{L}]=\mathbf{E}[Y_{1}^{\theta_1}]+\sum\limits_{l=2}^{L}\mathbf{E}[Y_{l}^{\theta_l}-Y_{l-1}^{\theta_l}],~\text{where}~Y_{l}^{\theta}= G(X_{l}^{\theta})e^{-\innerproduct{\theta_l}{ W_{T}^l}-\frac{1}{2}\lvert\theta_l\rvert^{2}T}.
\end{equation}
Under the framework of multilevel estimator, the parametric importance sampling estimator looks like,
\begin{equation}
\label{rev:eq10}
\displaystyle\widehat{Y}_L^{\theta}=\frac{1}{N_{1}}\sum\limits_{k=1}^{N_{1}}Y_{1}^{k,\theta_1}+\sum\limits_{l=2}^{L}\frac{1}{N_{l}}
\sum\limits_{k=1}^{N_{l}}(Y_{l}^{k,\theta_l}-Y_{l-1}^{k,\theta_{l}}).    
\end{equation}
Considering the variance of the above estimator, we have \cite{kebaier2018coupling},
\begin{equation}
\label{rev:eq11}
\displaystyle\text{Var}[\widehat{Y}_L^{\theta}]=\frac{1}{N_{1}}\mathbf{v}_{1}(\theta_1)+\sum\limits_{l=2}^{L}\frac{1}{N_{l}}
\sum\limits_{k=1}^{N_{l}}\frac{(M-1)T}{M^l}\mathbf{v}_{l}(\theta_{l}),    
\end{equation}
where,
\[\mathbf{v}_{1}(\theta_{1})=\text{Var}[Y_{1}^{\theta_{1}}]~\text{and}~\mathbf{v}_{l}(\theta_{l})=\text{Var}[Y_{l}^{\theta_{1}}-Y_{l-1}^{\theta_{1}}].\]
Therefore, as discussed, in order to solve the problem of minimizing the overall variance of the estimator described above, we intend to minimize the variance at each level of resolution, \textit{i.e.,} we aim at approximating $\theta_{l}^{*}$ for $l=1,\dots,L$, such that,
\begin{equation}
\label{rev:eq12}
\theta_{1}^{*}=\arg\min_{\theta \in \mathbf{R}^{d}}v_{1}(\theta_{1})~\text{and}~\theta_{l}^{*}=\arg\min_{\theta \in \mathbf{R}^{d}}v_{l}(\theta_{l}).
\end{equation}
Further, pertinent to the discussion carried out in \cite{alaya2015importance} and another application of the Girsanov's Theorem,  the above problem can be reformulated as,
\begin{eqnarray}
\label{rev:eq13}
\theta_{1}^{*}&=&\arg\min_{\theta_{1}\in \mathbf{R}^{d}}\mathbf{E}\left[G(X_{1})^{2}e^{-\innerproduct{\theta_1}{W_{T}^{1}}+\frac{1}{2}\lvert\theta_{1}\rvert^{2}T}\right] \nonumber\\
\theta_{l}^{*}&=&\arg\min_{\theta \in \mathbf{R}^{d}}\mathbf{E}\left[\frac{M^{l}}{(M-1)T}\left(G(X_{l})-G(X_{l-1})\right)^{2}
e^{-\innerproduct{\theta_l}{W_{T}^{l}}+\frac{1}{2}\lvert\theta_{l}\rvert^{2}T}\right].
\end{eqnarray}
We present below the two algorithm namely, the sample average approximation and stochastic approximation, in order to approximate the $\theta_{l}$'s as the solution to \eqref{rev:eq13}.
\subsection{Sample Average Approximation}
The sample average approximation deals with  approximating the above expectations using $\widetilde{N}_l$ sample paths. More specifically,
\begin{equation}
\label{rev:eq14}
\mathbf{E}\left[G(X_{1})^{2}e^{-\innerproduct{\theta_1}{ W_{T}^{1}}+\frac{1}{2}\lvert\theta_{1}\rvert^{2}T}\right] \approx \frac{1}{\widetilde{N}_{1}}\sum\limits_{j=1}^{\widetilde{N}_{1}}G(X_{1}^{k})^{2} e^{-\innerproduct{\theta_{1}}{W_{T}^{1,k}}+\frac{1}{2}\lvert\theta_{1}
\rvert^{2}T} \equiv \mathcal{V}_{1},
\end{equation}
and,
\begin{eqnarray}
\label{rev:eq15}
&& \mathbf{E}\left[\left(G(X_{l})- G(X_{l-1})\right)^{2}e^{-\innerproduct{\theta_l}{W_{T}^{l}}+
\frac{1}{2}\lvert\theta_{l}\rvert^{2}T}\right] \nonumber\\
& \approx & \frac{1}{\widetilde{N}_{l}}\sum\limits_{j=1}^{\widetilde{N}_{l}}\frac{M^{l}}{(M-1)T}\left(G(X_{l}^{k})-G(X_{l-1}^{k})\right)^{2}e^{-\innerproduct{\theta_{l}}{ W_{T}^{l,k}}+\frac{1}{2}\lvert\theta_{l}\rvert^{2}T} \equiv \mathcal{V}_{l}.
\end{eqnarray}
Having approximated the expectation in the minimization problem, the authors used the standard Newton-Raphson algorithm on the functions $\mathcal{V}_1$ and $\mathcal{V}_l$ in order to approximate $\theta_{l}^{*}$ for $l=1,\dots,L$. In \cite{alaya2015importance} it is proved that if the functional $G(X)$ satisfies the non-degeneracy conditions \textit{i.e.,} $\mathbf{P}((G(X^{1}_{T})\neq 0) >0 $ and  
$\mathbf{P}\left(((G(X^{l}_{T})-G(X^{l-1}_{T}))\neq 0\right) >0 $ and further assuming they have finite second moment, then by Lemma 2.1 in  \cite{alaya2015importance}, $\mathcal{V}_{1}$ and $\mathcal{V}_{l}$ are infinitely continuously differentiable. Moreover, both $\mathcal{V}_{1}$ and $\mathcal{V}_{l}$ are both strongly convex, thus implying the existence of the unique minimum $\theta_{1}^{*}$ and $\theta_{l}^{*}$ as the solution to equation \eqref{rev:eq13}.  

\subsection{Adaptive Stochastic Approximation}

Under the stochastic approximation, studied in \cite{alaya2016improved} the aim of determining the optimal change of parameter $\theta_{l}^{*}$ for $l=1,\dots,L$ is carried out using the Robbins-Monro algorithm. Here, we briefly describe the algorithm. Consider a compact convex set $\Theta \subset \mathbb{R}^{q}$ such that $ 0 \in \text{int}(\Theta)$. Then the recursive algorithm with projection is defined as follows,
\begin{equation}
\label{rev:eq16}
\theta_{l}^{n+1}=\textbf{Proj}_{\Theta}\left[\theta_{l}^{n}-\gamma_{n+1}H_{l}(\theta_{l}^{n},Y_{l},W_{T}^{l})\right],
\end{equation}
where,
$\text{Proj}_{\Theta}(\theta)=\min_{\theta \in \Theta}\abs{\theta-\theta_{0}}$.
The sequence $(\gamma_n)_{n \geq 1}$ is a decreasing sequence of positive real numbers satisfying,
\begin{equation}
\label{rev:eq17}
\sum\limits_{n = 1}^{\infty} \gamma_{n}=\infty~\text{and}~\sum\limits_{i=1}^{\infty} \gamma_{n}^{2} < \infty.
\end{equation}
Also,
\begin{equation}
\label{rev:eq18}
H_{l}(\theta_{l}^{n},Y_{l},W_{T}^{l})=\left\{\begin{array}{ll}
\left(\theta_{1}T-W_{T}^{1}\right)\left(G(X_{1})^{2}e^{-\innerproduct{\theta_1}{W_{T}^1}+\frac{1}{2}\lvert\theta_{1}\rvert^{2}T)}\right),&l=1, \\
\left(\theta_{l}T-W_{T}^{l}\right)\left[\frac{M^{l}}{(M-1)T}
\left(G(X_{l})-G(X_{l-1})\right)^2e^{-\innerproduct{\theta_l}{W_{T}^{l}}+\frac{1}{2}\lvert\theta_{l}\rvert^{2}T}\right],& l=2,\dots,L.
\end{array} \right\} 
\end{equation}
The algorithm described above is the constrained version of the Robbins-Monro algorithm. The inclusion of the projection operator in the recursive algorithm is to satisfy the non-explosion condition described above. Similar to the discussion carried out in the previous section, if the non-degeneracy conditions are satisfied \textit{i.e.,} $\mathbf{P}\left(G(X^{1}_{T})\neq 0\right) >0 $ and $\mathbf{P}\left(\left(G(X^{l}_{T})-G(X^{l-1}_{T})\right)\neq 0\right) >0 $, further assuming the finite second moment of $G(X_{1})$ and $G(X_{l})- G(X_{l-1})$, we can conclude the convergence of the $\theta_{l}^{*}$, constructed recursively using equation \eqref{rev:eq16}, for various level of resolutions.

The term adaptive is used in the sense that, the estimation of the optimal importance sampling parameter and the multilevel Monte Carlo run simultaneously. The multilevel estimator in this case is given as follows,
\begin{equation}
\label{rev:eq19}
\widehat{Y}_{L}^{\theta}=\frac{1}{N_{1}}\sum\limits_{k=1}^{N_{1}}Y_{1}^{k,\theta_{1}^{k-1}}+\sum\limits_{l=2}^{L}\frac{1}{N_{l}}\sum\limits_{k=1}^{N_{l}}\left(Y_{l}^{k,\theta_{l}^{k-1}}-Y_{l-1}^{k,\theta_l^{k-1}}\right).
\end{equation}
However, for the purpose of the practical implementation, one needs to stop the approximations procedure after finite number of iterations.

Having approximated the $\theta_{l}^{*}$ for $l=1,\dots,L$, we use the multilevel algorithm described by equation \eqref{rev:eq16} to estimate our expectation. It is quite evident from the way the algorithms have been described that the importance sampling algorithm combined with a multilevel estimator is more computationally complex than the standard multilevel algorithm. However, the variance reduction achieved by these combinations compensates for the high computational complexity. That is, the hybrid algorithm achieves the desired RMS error much faster than the MLMC estimator. The studies carried out in \cite{kebaier2018coupling,alaya2015importance} demonstrate the accuracy of the hybrid importance sampling multilevel algorithm over standard multilevel algorithm, through a series of numerical examples, where the underlying SDEs are multi-dimensional. The slight drawback of the sample average approximation method, though more stable than the adaptive stochastic algorithm, is the slow convergence rate to the optimal value. As for the stochastic approximation, the algorithm is sensitive to the learning parameter $\gamma_n$ and therefore is unstable. It may be pointed out that the study performed above only deals with the Euler Multilevel Monte Carlo, restricted to the use of Euler discretization to simulate the underlying SDEs. More recently, a study carried out by authors in \cite{sinha2022multilevel} generalize this approach, undertaking higher order discretization schemes such as Milstein to simulate the underlying SDEs. The interested reader can refer to the references mentioned therein to get a more rigorous understanding of this hybrid algorithm.

\section{MLMC and Efficient Risk Estimation.}
\label{multi_risk}

Risk measurement and consequent management is one of the essential components of financial engineering. The computation of the former (risk measures) for a financial portfolio is both challenging and computationally intensive, which may be ascribed to computations involving nested expectation, which entails multiple evaluations of the loss to the portfolio, for distinct risk scenarios. Further, the cost of computing loss of portfolio entailing thousands of derivatives
becomes progressively expensive with an increase in the size of the portfolio \cite{giles2019sub}. Value-at-Risk (VaR), Conditional VaR (CVaR), and the likelihood of a large loss are the necessary risk metrics used to estimate the risk of a financial portfolio. At the core of these estimation, is the necessity of evaluating the nested expectation, given by,
\begin{equation}
\label{rev:eq20}
\eta=\mathbf{E}\left[H\left(\mathbf{E}[X\lvert Y]\right)\right ]
\end{equation}
where, $H$ is the Heaviside function. More specifically, suppose we need to compute the probability of the expected loss being greater than $L_{\eta} \in \mathbf{R}$, \textit{i.e.,} we are interested in the following computation:
\begin{equation}
\label{rev:eq21}
\eta=\mathbf{E}\left[H(\mathbf{E}[\Delta\lvert R_{\tau}]-L_{\eta})\right],
\end{equation}
where $\mathbf{E}[\Delta \lvert R_{\tau}]$ is the expected loss in a risk-neutral world, with $R_{\tau}$ being a possible risk scenario at some short risk (time) horizon $\tau$. Also, $\Delta$ is the average loss of many losses incurred from different financial derivatives, depending upon similar underlying assets \cite{giles2019sub}, that is,
\begin{equation}
\label{rev:eq22}
\Delta=\frac{1}{K}\sum\limits_{i=1}^{K} \Delta_{i},
\end{equation}
where $K$ is the total number of derivatives and $\Delta_i$ is the loss from the $i$-th derivative. The average is considered to ensure the boundedness of $\Delta$, when the portfolio size of $K$ increases. A standard and straight forward way to estimate the nested expectation  \eqref{rev:eq20} is the usage of Monte Carlo method. This involves, simulating $M$ independent scenarios of the risk parameter $R_{\tau}$, and for each risk scenario, $N$ total
loss samples, which are independent. This method was explored in \cite{gordy2010nested} and an extended analysis was carried out in \cite{giorgi2017limit}. The total computational cost to perform the above simulation is $O(\max(K\epsilon^{-2},\epsilon^{-3}))$ in order to achieve the root-mean-squared (RMS) error of $\epsilon$ \cite{giles2019sub}. In order to handle this issue we present the ideas studied in \cite{giles2019multilevel} under the realm of MLMC.

\subsection{Adaptive Sampling Multilevel estimator}
\label{adap_multi}

As mentioned in the previous section, the cost of the standard Monte Carlo to achieve the root-mean-squared error of $\epsilon$ is $O(\epsilon^{-3})$. To improve the computational complexity, the authors in \cite{broadie2011efficient} developed an efficient through the adaptation of the sample size required in the inner sampler of Monte Carlo, to the particular outer sampler random variable. Under certain conditions, the authors were able to achieve the $O(\epsilon^{-5/2})$ computational complexity to achieve the RMS of $\epsilon$. Giles in \cite{giles2019multilevel} extended this approach to the multilevel framework and was able to achieve $O\left(\epsilon^{-2}\lvert \log \epsilon \rvert^{2}\right)$ computational cost for a RMS error tolerance $\epsilon$. Before presenting the work initiated by Giles, we put forth a brief review of the studies carried out in \cite{gordy2010nested} and \cite{broadie2011efficient}.

The authors in \cite{gordy2010nested}, estimated the inner expectation of the equation \eqref{rev:eq20}, \textit{i.e.,} $\mathbf{E}[X\lvert Y = y]$, for a given $y$, using the unbiased Monte Carlo estimator, with $N$ sample paths, as given by,
\begin{equation}
\label{rev:eq23}
\widehat{Z}_{N}(y)=\frac{1}{N}\sum\limits_{n=1}^{N} x_{n}(y),
\end{equation}
where, $\{x_{n}(y)\}_{n}$ are the mutually independent samples from the random variable $X$, conditioned on $Y=y$. Again using the Monte Carlo for the outer expectation, we have, 
\begin{equation}
\label{rev:eq24}
\eta \approx \frac{1}{M}\sum\limits_{m=1}^{M} H\left(\widehat{Z}_{N}(y_{m})\right), 
\end{equation}
where $\{y_{m}\}_{m}$ are the mutually independent samples from the random variable $Y$. Further, they proved that if the two random variables $\mathbf{E}[X\lvert Y]$ and $\widehat{Z}_N$ have the joint density $d_{N}(y,z)$ and assuming that for $i=0,1,2$, $\displaystyle{\frac{\partial}{\partial y_{i}}d_{N}(y,z)}$ exists, plus there exists a non-negative function $d_{i,N}$, such that,
\[\left\lvert \frac{\partial}{\partial y_{i}}d_{N}(y,z) \right \rvert \leq d_{i,N},~\text{for all}~ N,y,z,~\text{and}~ \sup_{N} \int\limits_{-\infty}^{\infty} \lvert z \rvert^{q}d_{i,N}(z)dz < \infty \tag{$\text{C}_2$} \label{rev:assumption_2},\]
for all $0\leq q \leq 4$, then the RMS error of the estimator \eqref{rev:eq24} is $O\left(M^{-1/2}+N^{-1}\right)$. Therefore, in order to achieve the RMS error of $O(\epsilon)$ we need $M=O(\epsilon^{-2})$ and $N=O(\epsilon^{-1})$, leading to the total computational complexity of $O(\epsilon^{-3})$. Authors in \cite{broadie2011efficient} developed an adaptive sampling technique to deal with high computational complexity  previously discussed. Their approach was based on the likelihood that an additional sample will result in a negative estimate of $\widehat{Z}_{N+1}$ having estimated that $\widehat{Z}_{N} > 0$ for given $Y$. More specifically they showed that,
\[\mathbf{P}\left[\widehat{Z}_{N+1}\leq 0 \lvert \widehat{Z}_{N}\right] \leq \frac{\sigma^{2}}{\left(N\widehat{Z}_{N}(Y)+\mu\right)^{2}}\approx \frac{\sigma^{2}}{N^{2}\mu^{2}},\]
where $\mu = \mathbf{E}[X\lvert Y]$ and $\sigma^2 = \text{Var}[X\lvert Y]$. Therefore, if $\displaystyle{N \geq \frac{\epsilon^{-1/2}\sigma}{\abs{\mu}}}$, then the probability that $H\left(\widehat{Z}_{N}(Y)\right)= H\left(\widehat{Z}_{N+1}(Y)\right) \approx H\left(\mathbf{E}[X \lvert Y]\right)$ is equal to $1-\epsilon$. Based on these observations, the authors in \cite{broadie2011efficient} introduced two algorithms, the first being based on the minimization of the total number of samples for all inner Monte Carlo samplers with respect to given tolerance $\epsilon$, and the second being iterative, estimating $\abs{\mu}$ and $\sigma$ after every iteration, for given value of $Y$, using $N$ samples further adding more inner samples till $\displaystyle{\frac{N\mu}{\sigma}}$ exceeds some error margin threshold. Under these two algorithms it was observed that the overall computational complexity is $O(\epsilon^{-5/2})$ \cite{giles2019multilevel}. The authors in \cite{giles2019multilevel} introduced the above algorithms in the realm of multilevel simulation, wherein they used multilevel estimator in order to achieve an approximation to the outer expectation, while making use of the sample size in the inner expectation as the discretization parameter. More specifically,
\begin{equation}
\label{rev:eq25}
\widetilde{\eta} \coloneqq \sum\limits_{l=0}^{L}\frac{1}{M_{l}}\sum\limits_{m=1}^{M_{l}} H\left( \widehat{Z}_{N_{l}}^{f,l,m}(y^{l,m})\right)-H\left(\widehat{Z}_{N_{l-1}}^{c,l,m}(y^{l,m})\right),
\end{equation}
where,
\begin{equation}
\label{rev:eq26}
\widehat{Z}_{N_{l}}^{f,l,m}(y)=\frac{1}{N_l}\sum\limits_{n=1}^{N_{l}} x^{f,l,m,n}(y),
\end{equation}
with $\{x^{.,l,m,n}(y)\}$ being the i.i.d samples of the random variable $X$, given $Y=y$. Also, $H\left(\widehat{Z}_{-1}^{c,0,\dots}(y)\right)\equiv 0$. Now under the assumptions \ref{rev:assumption_2}, it can be proved that \cite{giles2019multilevel},
\[\babs{\mathbf{E}\left[H\left(\widehat{Z}_{N_{l}}(Y)\right)-H\left(\mathbf{E}[X\lvert Y]\right)\right]}=O\left(N_{l}^{-1}\right).\] Further, under the assumption that there exists constants $\delta_{0}$ and $\rho_{0}$ such that, $\displaystyle{\rho(\delta)\leq \rho_{0}}$, for all $\delta \in [0,\delta_0]$ where $\delta$ is the random variable with density $\rho$, the authors in \cite{giles2019multilevel} proved that, if $X$ and $Y$  are the two random variables, satisfying the stated assumption, then,
\begin{equation}
\label{rev:eq27}
\text{Var}\left[H\left(\widehat{Z}_{N}(Y)\right)-H\left(\mathbf{E}[X\lvert Y]\right)\right]=O(N^{-1/2}).
\end{equation}
The above result determines the strong convergence property necessary to analyze the full potential of the MLMC estimator, in this scenario. However, if $\displaystyle{N_{l}=N_{0}2^{l}}$, then with standard MLMC complexity analysis it is easy to determine that the computational complexity required to achieve RMS error of $\epsilon$, we need $O\left(\epsilon^{-5/2}\right)$ computational complexity. To cater to this high computational demand, even in the framework of MLMC, the authors undertook the adaptive approach developed in \cite{broadie2011efficient} and extended it to the framework of MLMC.

Giles extended the studies carried out by authors in \cite{broadie2011efficient} to multilevel paradigm with an aim to reduce the overall computational cost to $O\left(\epsilon^{-2}\abs{\log \epsilon)}^2\right)$. In addition to the assumptions stated above, it is further assumed that,
\begin{equation}
\label{assumption_3}
\sup_{y} \mathbf{E}[\sigma^{-q}\abs{X-\mathbf{E}[X\lvert Y]}^{q}\lvert Y=y] < \infty, \quad 2 < q < \infty. \tag{$C_3$} 
\end{equation}
Thus, under the above stated assumptions, it was proved in Lemma 2.5 (for the perfect adaptive sampling) and Theorem 2.7 of \cite{giles2019multilevel}, that if the maximum number of sample paths is restricted to, 
\begin{equation}
\label{rev:eq28}
N =\bigg \lceil\max\left(O\left(\epsilon^{-1}\right),C^{2}\frac{\sigma^2}{\abs{\mu}^2}\right)\bigg \rceil,
\end{equation} 
then the further number of sample path of various level of resolutions are given by,
\begin{equation}
\label{rev:eq29}
N_{l}=\bigg \lceil N_{0}4^{l}\max\left(2^{-l},\min\left(1,\left(C^{-1}
N_{0}^{1/2}2^{l}\frac{\abs{\mu}}{\sigma}\right)^{-r}\right)\right)\bigg \rceil,
\end{equation}
with $C$ being some confidentiality constant and $\displaystyle{1 < r < 2-\frac{2}{q}}$ for the perfect adaptive sampling and \\ $\displaystyle{1< r < 2-\frac{\sqrt{4q+1}-1}{q}}$ when the values of $\abs{\mu}$ and $\sigma$ is approximated. Therefore,
\begin{equation}
\label{rev:eq30}
\text{Var}\left[H\left(\widehat{Z}_{N}(Y)\right)-H\left(\mathbf{E}[X\lvert Y]\right)\right]=O\left(2^{-l}\right),
\end{equation}
thereby leading to the overall computational complexity of the desired order. In a detailed discussion carried out in Section 4 of \cite{giles2019multilevel}, it was proved (pertaining to the calculation of VaR and CVaR) that in order to achieve the overall computational cost of $O(\epsilon)$ RMS error, the required computational complexity is $O\left(\max(\epsilon^{-2}\abs{\log \epsilon)}, K \epsilon^{-2})\right)$ for the estimation of VaR and CVaR, respectively. The numerical test on a model problem undertaken shows the efficacy of the algorithm constructed. Readers are directed to the referred paper for detailed discussion on the proofs of the above stated results. It may be noted that the computational complexity increases with an increase in the portfolio size, $K$. A random sub-sampling approach, extending it to a multilevel framework, thereby addressing the dependency on the portfolio size, to achieve the desired RMS error was recently introduced in \cite{giles2019sub}.

\section{Conclusion}

In this paper, we gave a brief overview of the recent trends in the paradigm of the multilevel algorithm concerning the importance sampling, in the case of option pricing and an adaptive sampling approach while determining the VaR and CVaR for large portfolios. The algorithms discussed serves as the improvement in the computational efficiency of the standard multilevel estimators, each having its merits and shortcomings. As discussed in Section \ref{importance_sampling}, the importance sampling algorithm combined with multilevel estimators significantly decreases variance at various resolution levels. However, the decrease in variance comes at the cost of increased computational complexity in either case and an increase in the sensitivity to approximate the optimal parameter. As for developing the MLMC based algorithm for efficient risk estimation discussed in Section \ref{multi_risk}, the adaptive sampling approach introduced in this paradigm leads to a significant improvement in the overall computational complexity to achieve the desired root mean squared error. However, the dependence of computational complexity on the size of the portfolio is a subtle shortcoming of the discussed algorithm. Overall, the presented ideas have substantially contributed to the research and development of the multilevel algorithm for various applications encountered in financial engineering problems. However, the scope to enrich the standard algorithm with non-standard variance reduction techniques is still an exciting path for future research.\\ \\

%\textbf{CRediT author statement}\\
%\textit{\textbf{Devang Sinha}: Conceptualization, Writing- Original draft preparation.}\\
%\textit{\textbf{Siddhartha P. Chakrabarty}: Conceptualization, Writing- Reviewing and Editing.}\\
%\\
%\textbf{Acknowledgments}\\
%This research did not receive any specific grant from funding agencies in the public, commercial, or not-for-profit sectors.\\
%\\
%\textbf{Declaration of interests}
%
%\begin{todolist}
%\item The authors declare that they have no known competing financial interests or personal relationships that could have appeared to influence the work reported in this paper.
%\item The authors declare the following financial interests/personal relationships which may be considered as potential competing interests:\\ \\
%Please declare any financial interests/personal relationships which may be considered as potential competing interests here.
%\end{todolist}

\bibliographystyle{elsarticle-num}

\bibliography{BIBLIO_Review}

\begin{thebibliography}{10}
\expandafter\ifx\csname url\endcsname\relax
  \def\url#1{\texttt{#1}}\fi
\expandafter\ifx\csname urlprefix\endcsname\relax\def\urlprefix{URL }\fi
\expandafter\ifx\csname href\endcsname\relax
  \def\href#1#2{#2} \def\path#1{#1}\fi

\bibitem{giles2008multilevel}
M.~B. Giles, Multilevel monte carlo path simulation, Operations research 56~(3)
  (2008) 607--617.

\bibitem{giles2008improved}
M.~Giles, Improved multilevel monte carlo convergence using the milstein
  scheme, in: Monte Carlo and Quasi-Monte Carlo Methods 2006, Springer, 2008,
  pp. 343--358.

\bibitem{giles2013antithetic}
M.~B. Giles, L.~Szpruch, Antithetic multilevel monte carlo estimation for
  multidimensional sdes, in: Monte Carlo and Quasi-Monte Carlo Methods 2012,
  Springer, 2013, pp. 367--384.

\bibitem{giles2013multilevel}
M.~Giles, L.~Szpruch, Multilevel monte carlo methods for applications in
  finance, Recent Developments in Computational Finance: Foundations,
  Algorithms and Applications (2013) 3--47.

\bibitem{lemaire2017multilevel}
V.~Lemaire, G.~Pag{\`e}s, Multilevel richardson--romberg extrapolation,
  Bernoulli 23~(4A) (2017) 2643--2692.

\bibitem{belomestny2013multilevel}
D.~Belomestny, J.~Schoenmakers, F.~Dickmann, Multilevel dual approach for
  pricing american style derivatives, Finance and Stochastics 17~(4) (2013)
  717--742.

\bibitem{heinrich2001multilevel}
S.~Heinrich, Multilevel monte carlo methods, in: International Conference on
  Large-Scale Scientific Computing, Springer, 2001, pp. 58--67.

\bibitem{kebaier2018coupling}
A.~Kebaier, J.~Lelong, Coupling importance sampling and multilevel monte carlo
  using sample average approximation, Methodology and Computing in Applied
  Probability 20~(2) (2018) 611--641.

\bibitem{alaya2016improved}
M.~B. Alaya, K.~Hajji, A.~Kebaier, Improved adaptive multilevel monte carlo and
  applications to finance, arXiv preprint arXiv:1603.02959 (2016).

\bibitem{giles2019multilevel}
M.~B. Giles, A.-L. Haji-Ali, Multilevel nested simulation for efficient risk
  estimation, SIAM/ASA Journal on Uncertainty Quantification 7~(2) (2019)
  497--525.

\bibitem{giles2014antithetic}
M.~B. Giles, L.~Szpruch, Antithetic multilevel monte carlo estimation for
  multi-dimensional sdes without l{\'e}vy area simulation, The Annals of
  Applied Probability 24~(4) (2014) 1585--1620.

\bibitem{arouna2004adaptative}
B.~Arouna, Adaptative monte carlo method, a variance reduction technique
  (2004).

\bibitem{alaya2015importance}
M.~B. Alaya, K.~Hajji, A.~Kebaier, Importance sampling and statistical romberg
  method, Bernoulli 21~(4) (2015) 1947--1983.

\bibitem{chen1987convergence}
H.-F. Chen, L.~Guo, A.-J. Gao, Convergence and robustness of the robbins-monro
  algorithm truncated at randomly varying bounds, Stochastic Processes and
  their Applications 27 (1987) 217--231.

\bibitem{chen1986stochastic}
H.F.Chen, Y.~Zhu, Stochastic approximation procedures with randomly varying
  truncations, Science in China, Ser. A (1986).

\bibitem{andrieu2005stability}
C.~Andrieu, {\'E}.~Moulines, P.~Priouret, Stability of stochastic approximation
  under verifiable conditions, SIAM Journal on control and optimization 44~(1)
  (2005) 283--312.

\bibitem{lelong2008almost}
J.~Lelong, Almost sure convergence of randomly truncated stochastic algorithms
  under verifiable conditions, Statistics \& Probability Letters 78~(16) (2008)
  2632--2636.

\bibitem{lemaire2010unconstrained}
V.~Lemaire, G.~Pag{\`e}s, Unconstrained recursive importance sampling, The
  Annals of Applied Probability 20~(3) (2010) 1029--1067.

\bibitem{kloeden1992stochastic}
P.~E. Kloeden, E.~Platen, Stochastic differential equations, in: Numerical
  Solution of Stochastic Differential Equations, Springer, 1992, pp. 103--160.

\bibitem{sinha2022multilevel}
D.~Sinha, S.~P. Chakrabarty, Multilevel richardson-romberg and importance
  sampling in derivative pricing, arXiv preprint arXiv:2209.00821 (2022).

\bibitem{giles2019sub}
M.~B. Giles, A.-L. Haji-Ali, Sub-sampling and other considerations for
  efficient risk estimation in large portfolios, arXiv preprint
  arXiv:1912.05484 (2019).

\bibitem{gordy2010nested}
M.~B. Gordy, S.~Juneja, Nested simulation in portfolio risk measurement,
  Management Science 56~(10) (2010) 1833--1848.

\bibitem{giorgi2017limit}
D.~Giorgi, V.~Lemaire, G.~Pag{\`e}s, Limit theorems for weighted and regular
  multilevel estimators, Monte Carlo Methods and Applications 23~(1) (2017)
  43--70.

\bibitem{broadie2011efficient}
M.~Broadie, Y.~Du, C.~C. Moallemi, Efficient risk estimation via nested
  sequential simulation, Management Science 57~(6) (2011) 1172--1194.

\end{thebibliography}

\enlargethispage{1.4cm}

\end{document}